# Comparing the physical characteristics of ultrasound and magnetic resonance imaging to diagnose ovarian cysts

*Tariq Nadhim Jassim[1]* Radiological Techniques Department, College of Health and Medical Techniques, Al-Mustaqbal University, Hillah, Babylon, Iraq.
*Rasha Tahseen Ibrahim[2]* Medical Physics Department ,College of Science,Al-Ameen Unviersity,Baghdad,Iraq. (the authors would like to acknowledge the support of AL-Ameen University, Iraq for their valuable support)
*danadanadano8@gmail.com*
*Mohsen Hamoud Jasim[3]* - Medical Physics Department, Hilla University College, Babylon, Iraq.
*mohsen07829431518@gmail.com*
*NAHD JABBAR DALFI[4]* - College of Health and Medical Technologies - Baghdad
Radiology Technologies Department
*Mohammed Fatta[5]* - Radiological Techniques Department, College of Health and Medical Techniques, Al-Mustaqbal University, Hillah, Babylon, Iraq.
*Mohanad Ahmed Sahib[6]* -Al-Mustaqbal University, College of Health and Medical Techniques,Radiological Techniques Department ,Babylon, iraq

**Abstract:**
*Background,* For the purpose of determining the appropriate course of therapy to maintain fertility, a correct diagnosis of ovarian cysts is crucial.
*Objective*, To contrast the results of magnetic resonance imaging and ultrasonography in individuals with ovarian cysts
.

*Methods*: research was carried out in the radiology division of Al-Hilla General Teaching hospital and Marjan Teaching Hospital in  province Babil in the period of November 2023 and march 2024, Seventy-five Following a physical and ultrasonography examination, the female patient was assessed using magnetic resonance imaging.
Women in the over-18 age group exhibit a range of symptoms, such as irregular menstruation, abdominal pain, sensoria, and menorrhagia.
 *Results*: Adnexal lesions on US imaging included chocolate cysts (14.6%), dermoid cysts (13.8%), hemorrhagic cysts (HC) (11.6%), simple cysts (32.4%), complicated cysts (21.2%), and multilocular cysts (MC) (6.4%).Simple cysts (SC) made up 29%, complex cysts (6.9%), dermoid cysts (10.9%), chocolate cysts (19.1%), hemorrhagic cysts (HC) 11.3%, multilocular cysts (6.8%), and malignant cysts (MC) 16% of the adnexal lesions on MRI. The USG results show 100% sensitivity, 78.3% specificity, and 89.3% accuracy when compared to the MRI results.
 *Conclusions:* Magnetic resonance imaging, which is quite accurate in determining the mass's origin and characterizing its tissue content, may be beneficial in further evaluating monographically vague ovarian cysts with solid or complicated content.

**Keywords:** Ultrasound, magnetic resonance imaging , simple cyst, Haemorrhagic cysts, malignant cyst

## Introduction

An ovarian cyst is a fluid-filled sac that is visible from the time of birth until after menopause (1). Both functional and physiological cysts can result from ovarian cysts (produced by a follicle that fails to rupture) or corpus luteum cysts (made by hemorrhaging in a corpus

luteum)(2). Additional types of ovarian cysts include polycystic ovaries, endometrioid cysts (also known as chocolate cysts), dermoid cysts (teratomas), and cystadenomas. A majority of ovarian cysts are benign, harmless, might not show any symptoms, and can disappear on their own in a few months. If we perform a histological investigation, it could be challenging to distinguish benign from malignant ovarian cysts.(3). On the other hand, complications from the operation required to remove the cyst could include twisting, rupturing, bleeding, and pressure on the surrounding organs. Less than 1% of asymptomatic premenopausal women with unilocular ovarian cysts will develop ovarian cancer.
; however, premenarchal and postmenopausal women are more likely to develop cancer from these cysts(4,5). Simple ovarian cysts are more common than complex ones, but complex cysts carry a significant risk of cancer. Complicated cysts are filled with either blood or a coarse substance. Unlike simple cysts, specific cysts are not related to the regular menstrual cycle(6,7). Treatment for ovarian cysts is determined by the presentation (asymptomatic or cystic) as well as the likelihood of cancer. Most functioning cysts can be treated conservatively and disappear on their own in two cycles. Usually, when an ovarian cyst, presents with serious symptoms, medical intervention is necessary(8). Magnetic resonance imaging (MRI) and ultrasound (US) are the two main modalities frequently employed to identify cystic diseases(9). The predominant imaging method for adnexal mass classification and characterization is remains (US) (10). Since numerous facilities throughout the world have accumulated a wealth of knowledge, it is possible to appropriately classify about 90% of adnexal masses based on their US features(11). In order to assess which patients require surgery, what kind of surgery is needed, and whether a surgical subspecialist is needed, it is necessary to classify the adnexal mass (12).The importance of magnetic resonance imaging (MR imaging) in the assessment of patients with adnexal illness is evolving. Certain benign tumors, such as fibromas, hydrosalpinx, endometriomas, teratomas, and simple and hemorrhagic cysts, can be found with full information MR imaging .When it comes to lesion description, staging, and follow-up, MR imaging may be more helpful in patients with malignant lesions than other modalities. (13). Our study attempts to categorize the potential applications of magnetic resonance imaging (MRI) and ultrasound in the diagnosis of the most prevalent ovarian lesion.

**Method……**
**Study design and settings**

In the radiology departments of Al-Hilla General Teaching Hospital and Marjan Teaching Hospital in the province of Babil, 75 consecutive patients over the age of 18 were seen between November 2023 and March 2024. The patients' mean age was 39.11 years, and they were diagnosed with adnexal lesions by ultrasound and MRI. The women's symptoms varied and included irregular cycles, abdominal pain, bleeding, and incidental. Using a 1.5 T MR equipment, MR imaging was possible for T1-weight damages (T1W), T2-weighted pictures, and fat-suppression images. T1-weighted MRI contrast agent (gadolinium) images taken both before and after an IV. Shape, size, content (solid or cystic), nodal or vascular septum, and reinforcement are characteristics of ovarian cysts. Aside from that, ascites, enlarged lymph nodes, and peritoneal illness also manifested. We use the picture features to distinguish between surgical and pathologic findings. Multiple logistic regression analysis was done on all MR imaging features without requiring any clinical information. Based on the aspects of the images that distinguished them from the pathological findings and surgical results, they were categorized as either benign or malignant.

**ULTRASOUND APPARTUS:**

GE version E6, Philips HD11xe, and GE Vivid E9 transvaginal transducers measuring 6 MHZ were used to evaluate the patients; in certain cases, a curved array transducer measuring 2–5 MHZ was also used for transabdominal scanning. The patient was placed supine on a regular table, with their legs slightly bent and abducted, for the examination..

Because a bloated bladder might alter the anatomy of the pelvis, the patient must empty her bladder in order to improve the TVUS result. After a transverse and longitudinal examination of the uterus, the fallopian tube and ovaries in adnexa were assessed, and finally the CUL-DE-SAC.

**MRI APPARTUS:**

A 1.5-T system (Philips Ingenia and GE Optima –MR450W) was used to assess the patient. Using body coils and multi-coil arrays

**TECHNIQE:**

There was no need for preparation; the patient was simply positioned on their back, and the body coil was positioned on the pelvis so that its lower edge rested below the pubic bone. The patient is then moved head-first into the magnet's hole while the coil is secured in place using a belt.

Among the sequences are:
1. T2-weighted test-sagittal image: used to identify the uterine axis of mass.
2-Transverse weighted image (T2-tse): planed perpendicular to the mass axis as determined by the sagittal series.
To determine the link between mass, uterus, and adnexa, use 3-T2, test-coronal weighted picture. include the sides wall, pelvic floor, and ovaries.
4. T1-tse-transverse, both with and without fat suppression, is required to identify T1bright masses that contain fat.
5. Gadolinium should be administered whenever it is feasible. The contrast agent utilized was gadolinium chelates, which were manually administered intravenously using a cannula at a dose of 0.1 (mmol/kg).
Every scan was performed with a 1-mm gap and a 5-mm thickness, covering a 250–375 mm FOV.
the T1-weighted picture.T2 weighted image (TR4.0 sec / TE 100ms) is faster than TR (500ms/TE 20 ms).

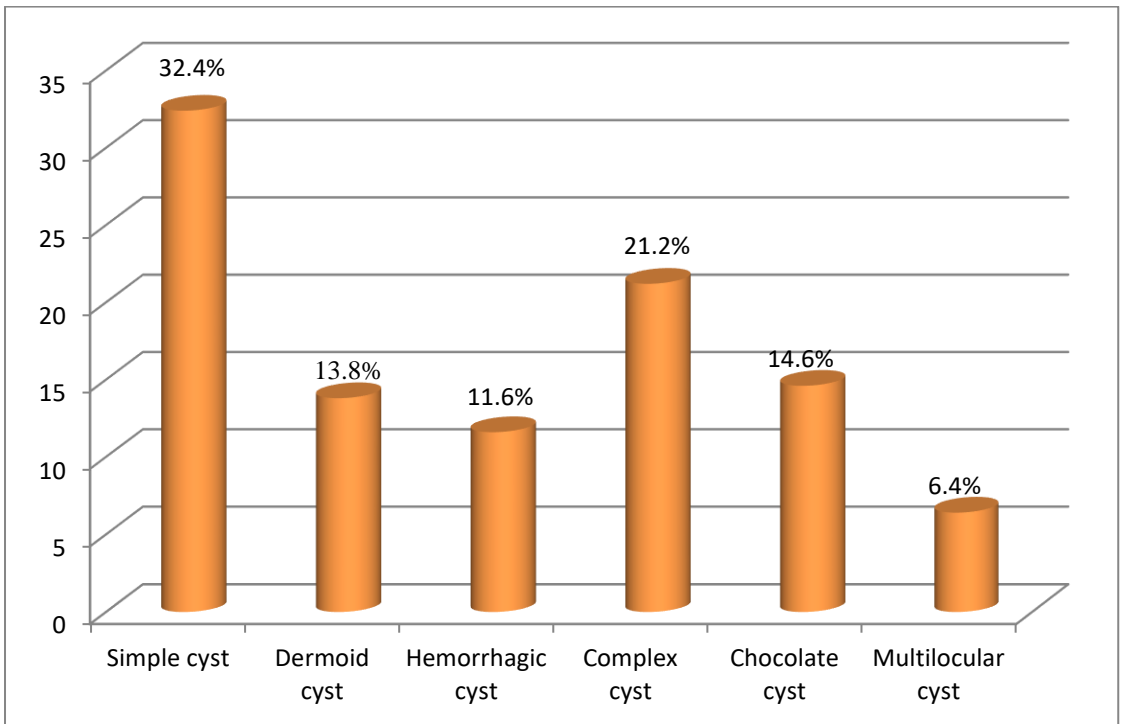

**Figure 1: explain** Patient distribution based on ovarian cyst ultrasound diagnosis

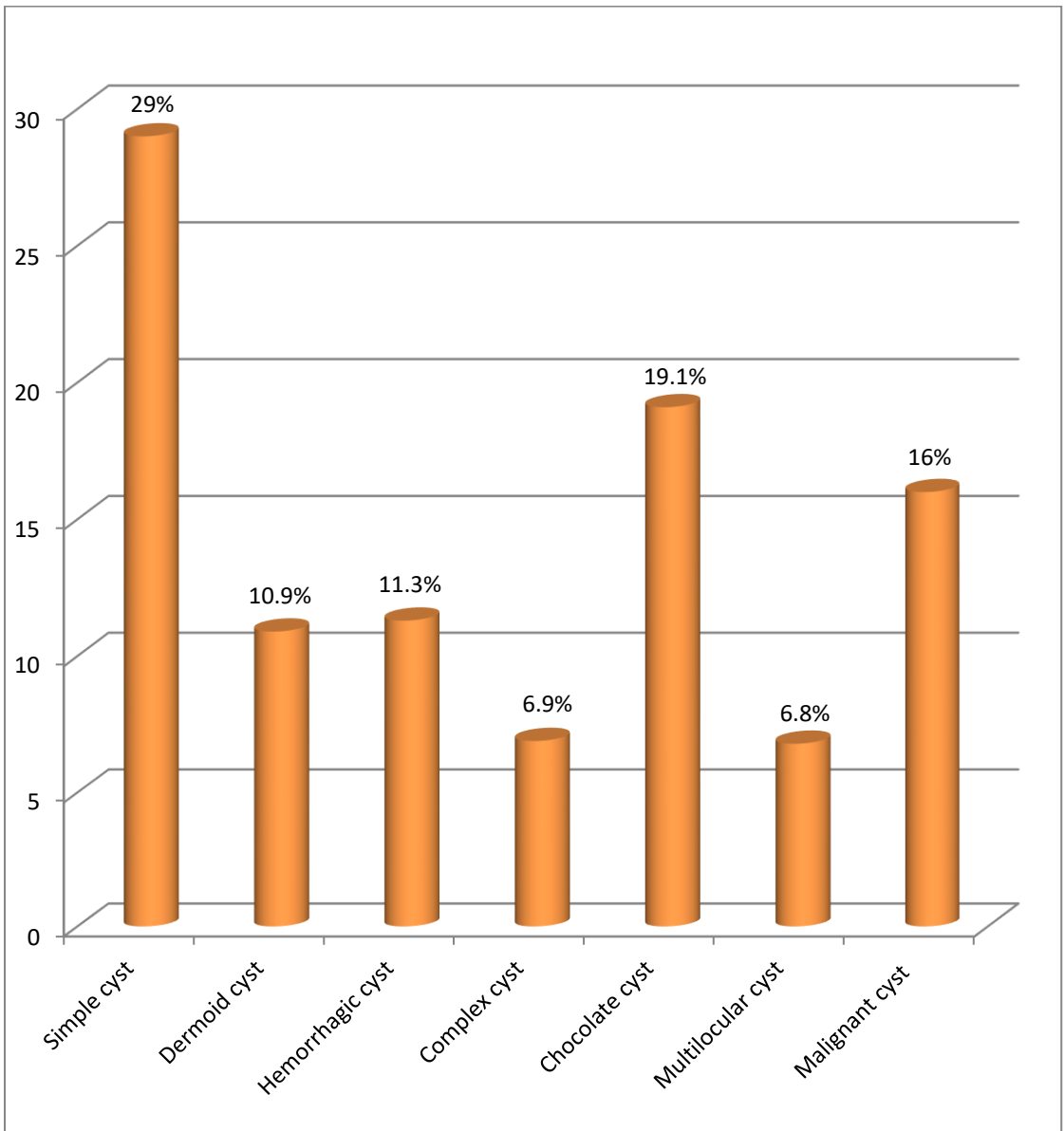

**Figure 2** :**explain** Patient distribution based on ovarian cyst MRI diagnostic

**Table (1) explain:** Comparing US and MRI data for the diagnosis of ovarian cysts in terms of sensitivity, specificity, negative predictive value, positive predictive value, and overall accuracy

| {Ultrasound diagnosis} | MRI diagnosis | | Total |
|---|---|---|---|
| | Malignant ovarian cyst | Benign ovarian cyst | |
| {Malignant ovarian cyst} | 12 | 8 | 20 |
| {Benign ovarian cyst} | 0 | 55 | 55 |
| {Total} | 12 | 63 | 75 |

US Sensitivity= 12/12 * 100= 100.0%
Particulars of the US= 55/63 * 100= 78.3%
(PPV) of US= 12/20 * 100= 57.14%

(NPV) of US= 55/55 * 100= 100.0%

The overall precision of the US= (12+55)/ 75 * 100= 89.3%

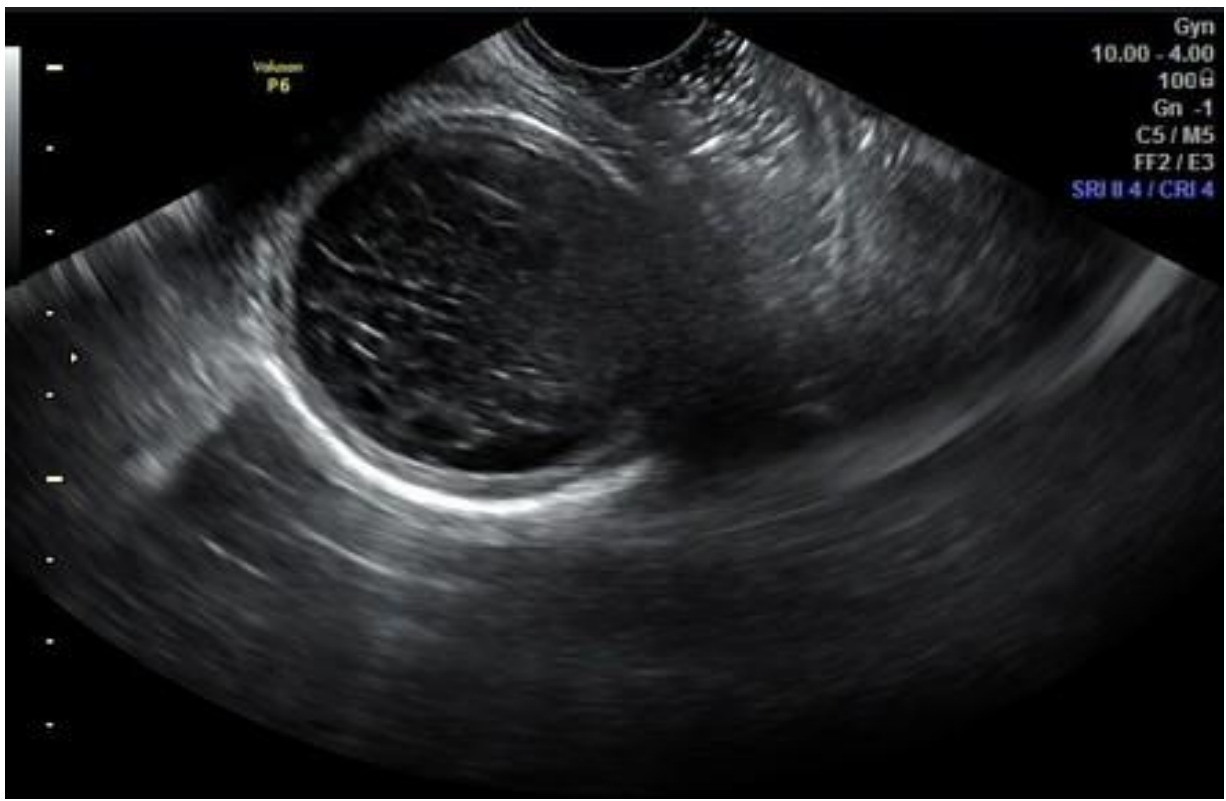

**Figure 3**: ultrasound for woman (35) years old presenting with recurrent right sided abdominal pain**,** showing hemorrhagic ovarian cyst in right ovary**.**

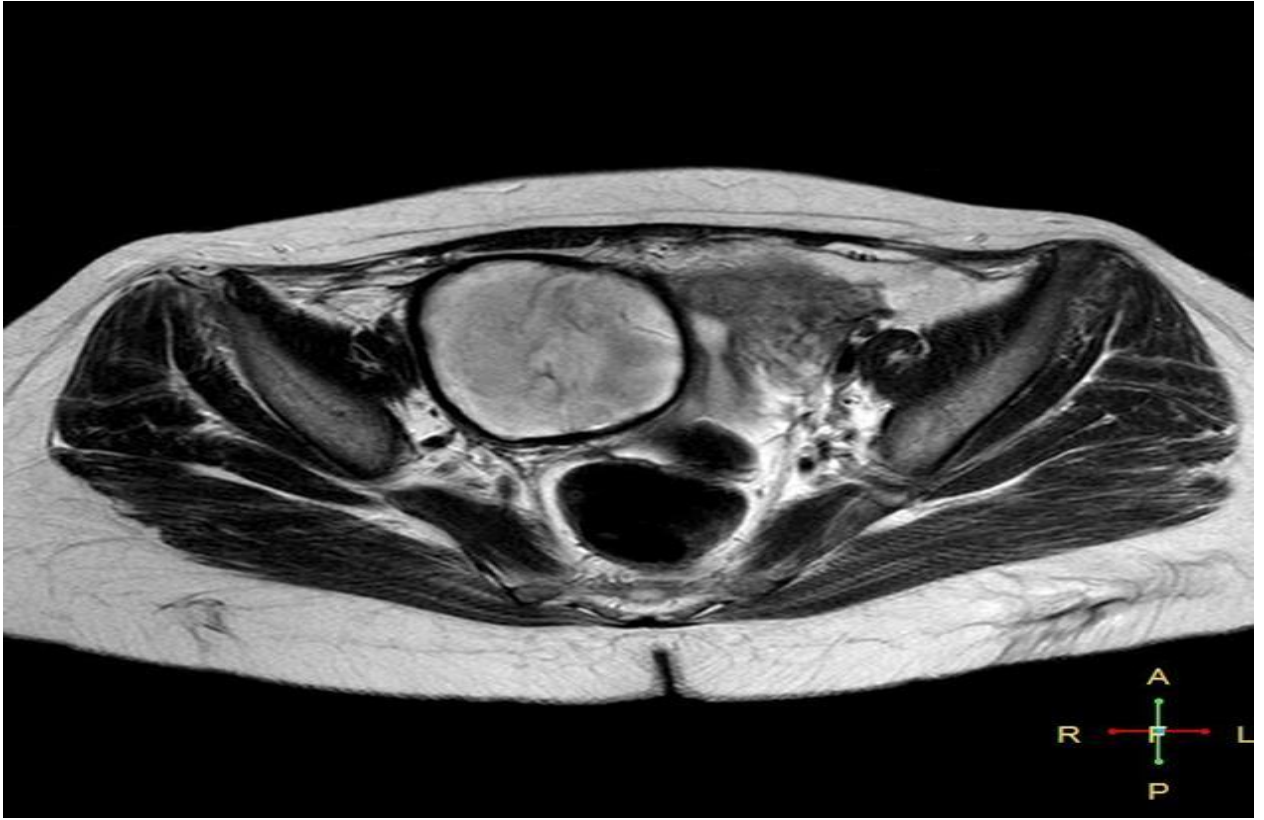

**Figure 4:** T2 weighted image, hyper intense signal with very low signal wall

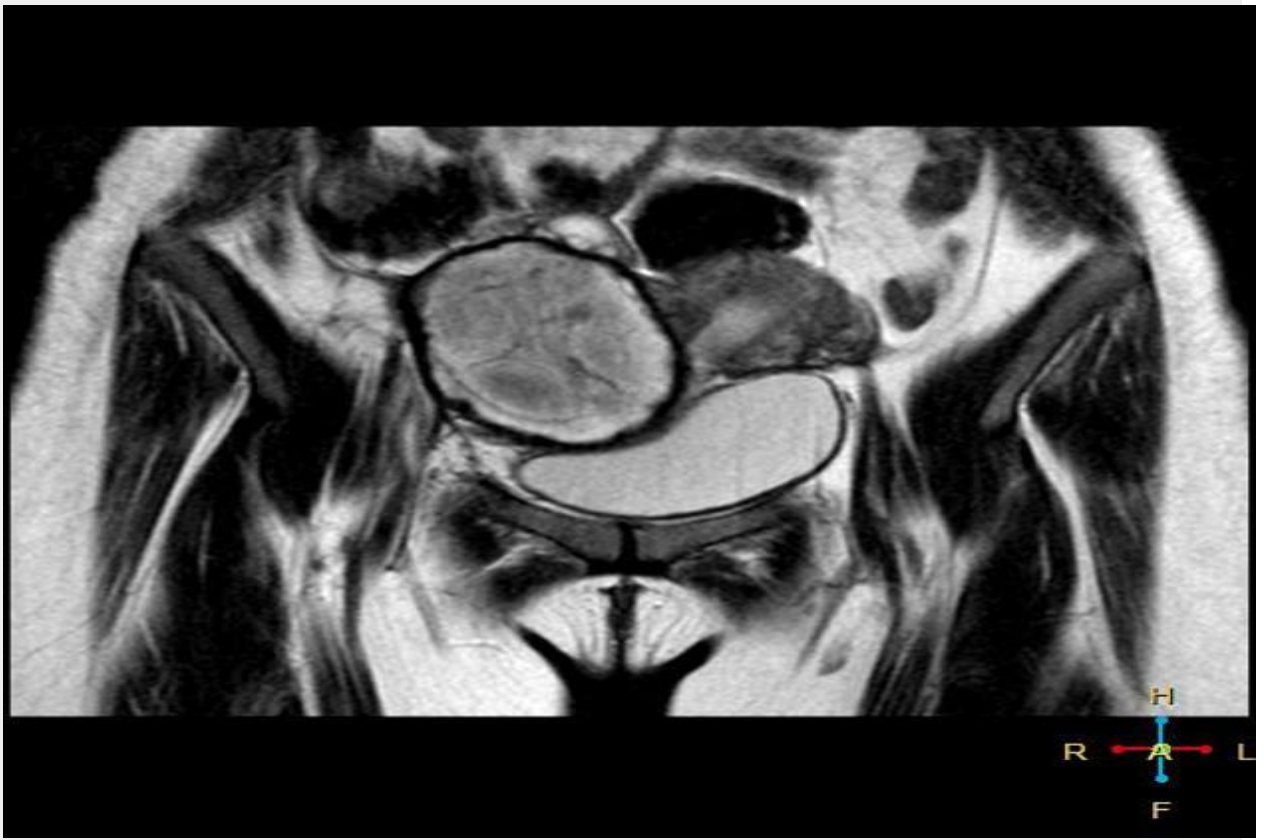

**Figure 6:** coronal T2 weighted image, hyper intense signal with very low signal wall

**Discussion**

The research involved 75 women with clinically suspected ovarian abnormalities, initially examined by ultrasound and subsequently by magnetic resonance imaging (MRI).
Patients using ultrasounds showed evidence of ovarian cancer. Simple cysts (32.4%), complicated cysts (21.2%), dermoid cysts (13.8%), chocolate cysts (14.6%), hemorrhagic cysts (11.6%), and multilocular cysts (6.4%) were the adnexal lesions.
75 women with ovarian lesions underwent magnetic resonance imaging. The adnexal lesions included malignant ovarian cysts (16%), hemorrhagic cysts (11.7%), multilocular cysts (6.8%), chocolate cysts (19.1%), simple cysts (29%) and complicated cysts (6.9%).
The US is more accurate in diagnosing ovarian cysts than MRI data in terms of sensitivity, specificity, positive predictive value, negative predictive value, and overall accuracy.
The accuracy of ultrasound in diagnosing malignant ovarian cysts was 100.0%, indicating that all patients with malignant ovarian cysts were appropriately diagnosed by ultrasound. The specificity of ultrasonography in the diagnosis of benign ovarian cysts was found to be 78.3%, meaning that 78.3% of patients were correctly identified with the condition.
Overall accuracy was (89.3%), with a positive predictive value of (57.14%) meaning that all patients diagnosed as benign ovarian cysts by US would also be diagnosed as benign ovarian cysts by MRI, and a negative predictive value of (100.0%) meaning that all patients diagnosed as malignant ovarian cysts by U/S would also be diagnosed as malignant ovarian cysts by MRI.
In this study, we found that a sonographer's ability to accurately diagnose adnexal lesions by ultrasound also depended on their proficiency and understanding of the characteristics of these lesions and how to differentiate between them.
Rather than sending the patient for an MRI, benign lesions such as simple cysts, hemorrhagic cysts, tiny Para ovarian cysts, small endometroid cysts, and hydrosalpinx should be monitored. MRIs should only be performed on individuals who are suspected of having cancer, cannot accurately determine the location of large tumors, and whose tumors do not appear to be shrinking on a follow-up scan. The validity results of this investigation are similar to those of the Ramya study (14).
The results of this investigation MRI is more useful for issue solving when US findings are inconclusive or non-diagnostic since, despite its higher cost, it is a more accurate diagnostic tool. A comprehensive approach to diagnosis is made possible by the signal intensity characteristics of ovarian masses. T1-weighted, T2-weighted, and fat-suppuration T1-weighted MRI imaging data can be used to correctly characterize mature cystic teratomas, endometriomas, leiomyomas, fibromas, and other lesions.

**Conclusion**

The capacity of magnetic resonance imaging (MRI) to precisely identify the source of a mass and describe its contents makes it a valuable tool in the assessment of adnexal masses. If sonographic tests are technically constrained, it would seem reasonable that individuals with a suspected adnexal mass might benefit from MRI.

According to our research, MRI further assessment might be beneficial if sonography reveals an adnexal mass with solid or complex cystic tissue, a big mass, or a doubtful pedunculated uterine fibroid vs ovarian tumor.